\documentclass[12pt]{article}
\usepackage{amsmath,amssymb,mathrsfs}
\usepackage{psfrag}
\usepackage{graphicx}
\usepackage{graphics}
\usepackage{epsfig}
\usepackage{bm}
\usepackage{color}
\usepackage{verbatim,color,ulem}

\newcommand{\beq}{\begin{equation}}
\newcommand{\eeq}{\end{equation}}
\newcommand{\beqa}{\begin{eqnarray}}
\newcommand{\eeqa}{\end{eqnarray}}

\begin{document}

\title{Gauge approach to superfluid density in underdoped cuprates}
\author{P.A. Marchetti$^{1,2}$ and G. Bighin$^{1,2}$}
\date{}

\maketitle

\begin{center}
$^{1}$Dipartimento di Fisica e Astronomia ``Galileo Galilei'', \\ 
Universit\`a di Padova, Via Marzolo 8, 35131 Padova, Italy \\
\vspace{8pt}
$^{2}$Istituto Nazionale di Fisica Nucleare, Sezione di Padova, \\
Via Marzolo 8, 35131 Padova, Italy
\end{center}

\begin{abstract}
We prove that a gauge approach based on a composite structure of the hole in hole-doped cuprates is able to capture analytically many features of the experimental data on superfluid density in  the moderate-underdoping to nearly-optimal doping region, including critical exponent, the Uemura relation and near universality of the normalized superfluid density.
\end{abstract}

%\pacs{71.10.Hf, 71.27.+a, 11.15.-q, 74.25.-q, 74.72.Gh}

\section{Introduction}

The superfluid density $\rho_s$ of hole-doped cuprates in the region of moderate underdoping to nearly optimal doping as a function of temperature exhibits features whose combination does not allow for a simple explanation. The well-defined gapped Fermi arcs seen in ARPES experiments\cite{leezxshen} and the linear $T$-dependence near $T=0$ of $\rho_s$ hints at a BCS-like hole dynamics.
Attributing the low temperature $T$-linear dependence to the quasi-particle excitations near the nodes of the d-wave BCS order parameter \cite{lee}, however, it is difficult to reconcile with the non-mean-field critical exponent $\approx 2/3$, typical of a 3DXY model, at the superconducting transition \cite{kamal}(in a BCS theory one would expect the mean-field value 1). Also it appears at odds   with the intriguing Uemura linear relation \cite{uemura} between $\rho_s(T=0)$ and $T_c$ approximately verified for moderate underdoping (in a BCS theory the superfluid density at $T=0$ does not depend on the BCS order parameter controlling $T_c$).
An alternative explanation of the $T$-linear behaviour is based on phase \cite{kivel} or pairing fluctuations \cite{levin} of the order parameter in a BCS-BEC crossover setting. In both cases one can obtain a low-energy effective theory in terms of the phase field described by an XY model, whose superfluid density exhibits the desired linear slope. Typically the advocated XY model is two-dimensional, but then the critical behaviour comes out experimentally incorrect. A 3-dimensional nature of the XY model is sometimes claimed to emerge in a narrow range of temperatures close to $T_c$ due to the presence of a stack of Cu-O layers in the high-$T_c$ materials. However the range of a 3-dimensional behaviour  found experimentally is larger than expected and when one crosses to the region where the 2-dimensional behaviour is anticipated one does not find the jump at the Kosterlitz-Nelson temperature, characteristic of superfluid density in the BKT formalism\cite{hardy}. Moreover, it is possible to interpret the large pseudogap experimentally observed above $T_c$ in terms of the BCS-BEC crossover, however it would be at odds with the presence of a well defined gapped Fermi surface.

In this paper we show that a solution to the above puzzles is provided by a gauge approach developed in \cite{msy}\cite{mfsy}\cite{mg}, based on a modelling of the low-energy physics of Cu-O planes by a planar $t-t'-J$ model with square lattice and heavily relying on a composite nature of the hole. Such model is widely believed to be appropriate for the description of hole-doped cuprates due to the large charge transfer gap in these materials, with Zhang-Rice singlets  as charge carriers, see e.g. \cite{damascelli}.
The hole in our  approach to this model is seen as a composite of a spin (spinon) and a charge (holon) excitation. As discussed below the specific choice adopted for these excitations is different from the standard slave-boson approach and it is suggested by 1D results.
Specifically the charge degrees of freedom are mainly responsible for the Fermi surface structure, whereas in the considered doping region the superfluid density is dominated by the spin excitations which are also controlling the superconducting transition. Since $T_c$ is not determined by the charge degrees of freedom, this dichotomy allows for the existence of a well defined Fermi surface along with both a large pseudogap region, needed for a pairing temperature for the charges well above the superconducting one, and with the strongly BEC behaviour, needed for the XY-like superfluid density and the Uemura relation. Furthermore, the spin excitation dynamics is ``relativistic" with space and time treated on the same footing and it is described by a 3D XY model with temperature dependent coupling constant, thus explaining the critical exponent of the superfluid density. Actually it turns out that, consistently with the experiments in the moderate underdoping to nearly optimal doping region,  the superfluid density normalized by its value at $T=0$ as a function of $T/T_c$ is well fitted by the 3DXY curve almost universally. Furthermore an approximate Uemura relation is analytically derived.

\section{The gauge approach}

We start sketching the key ideas of our gauge approach relevant to understand the treatment of the superfluid density. 
The no-double occupation of the the  $t-t'-J$ model is tackled with a spin-charge decomposition of the  hole $c$ written as a product of a spinon $s$, spin 1/2 boson, times a holon $h$, spinless fermion. Since $h$ is spinless, by Pauli principle the no-double occupation  constraint is automatically implemented. This decomposition introduces an unphysical degree of freedom due to a local $U(1)$ gauge invariance, because one can multiply at each site the spinon and the holon by arbitrary opposite phase factors leaving the physical hole field unchanged. This invariance is made manifest with the introduction of a slave-particle gauge field $A$, which in turn produces an attraction between $s$ and $h$. A long-wavelength treatment of the spinons leads to a CP$^1$ spinon nonlinear $\sigma-$model with $A$ as gauge field, producing a ``relativistic" dispersion relation for the spinons. In our approach we further use a feature specific to 2 (or 1) space dimensions: we add a 1/2 charge-flux $\Phi_h$  to $h$ and a 1/2 spin-flux $\Phi_s$  to $s$, still retaining the fermionic statistics of $c$ \cite{msy1}. 
%The gradients  $\nabla \Phi$  can be viewed as the potentials of  vortices.
 The spinless holons dressed by  $\Phi_h$ are expected to obey Haldane-Wu semion statistics  \cite{wu} allowing double occupation in momentum-space, as can be proved for their 1-dimensional analogue, where the statistics is crucial to get the correct critical exponents 
\cite{msy2}. We assume this property for our dressed holons \cite{igu} and
%. A gas of spinless semions of finite density has 
therefore these holons have a Fermi surface at low $T$ coinciding with that of spin 1/2 fermions of the same density. 
 The physical hole is obtained as a holon-spinon bound state produced by the gauge attraction and it inherits the features of the holon Fermi surface, but renormalized and strongly overdamped by the interaction of the spinon with gauge fluctuations.\cite{msy}
 
 We consider a mean-field (MF) approach where we neglect the holon fluctuation in $\Phi_h$ and the spinon fluctuations  in  $\Phi_s$. The result is for the spin-flux $\Phi_s(j)= (\sigma_z/2) \sum_i \arg(j- i)h_i^*h_i(-1)^{|i|}$ , where $j,i$ are sites of the lattice and $|i|=i_x+i_y$ .
The gradient  $\nabla \Phi_s$  can be viewed as the potential of $U(1)$ spin-vortices centered at holon positions with $(-1)^{|i|}$ chirality . Hence the empty sites of the model where holons sit are surrounded by spin-vortices, a quantum distortion of the antiferromagnetic (AF) spin background, with opposite chirality in the two N\'eel sublattices. These vortices are missed in the slave-boson treatment.
The interaction of spinons and spin-vortices in the long wavelength limit is of the form
\begin{equation}
\label{spvo}
J (1-2 \delta)(\nabla{\Phi^s(x)})^2  s^* s(x)
\end{equation}
and it is the source of both short-range AF and charge pairing. From a quenched treatment of spin-vortices we derive the MF expectation value
$\langle(\nabla{\Phi^s(x)})^2\rangle = m_s^2 \approx 0.5\delta |\log
\delta|$, which opens a mass gap for the spinon, consistent with AF
correlation length at small $\delta$ extracted from the neutron
experiments \cite{ke}.
 Thus, propagating in the gas of slowly moving spin-vortices, the AF spinons, originally gapless in the undoped Heisenberg model,
acquire a finite gap, leading to a short range AF order.
By averaging instead the spinons in Eq. (\ref{spvo}), we obtain an effective interaction:
 \begin{equation}
\label{zh}
J (1-2 \delta) \langle s^* s \rangle \sum_{i,j} (-1)^{|i|+|j|} \Delta^{-1}
 (i - j) h^*_ih_i h^*_jh_j,
\end{equation}
 where $\Delta$ is the 2D lattice Laplacian.
From Eq. (\ref{zh}) we see the interaction mediated by spin-vortices on holons is of 2D Coulomb type. From the known behaviour of planar Coulomb systems we derive that below a crossover temperature
$T_{ph} \approx J (1-2 \delta)  \langle s^* s \rangle $ (identifiable with $T^*$ in \cite{pnas}) a finite density of incoherent holon pairs appears.
Therefore the origin of the charge-pairing is magnetic since it involves the N\'eel chirality structure of vortices , but it is not due to exchange of AF spin fluctuations. 
In MF for the charge-flux we get $\Phi_h(j)=(1/2) \sum_i \arg(j- i)$ yielding a $\pi$ flux per plaquette converting the low-energy modes of the spinless holons $h$ into Dirac fermions with linear dispersion
% defined in the Magnetic Brillouin Zone (MBZ)
 and two small Fermi surfaces (FS) centered at $(\pm \pi/2, \pi/2)$ in the the magnetic Brillouin Zone (BZ) with Fermi energy $\epsilon_F \sim
t\delta$, characterizing what we name the ``pseudogap phase" (PG) of the model. By increasing doping or temperature one reaches a crossover line $T^* \leq T_{ph}$, identified with the experimental inflection point of in-plane resistivity (and with $T^{**}$ in \cite{pnas}), then entering the ``strange metal phase" (SM)  in which  the effect of the charge flux is screened by spinon phases and we recover a ``large'' FS for the holons with $\epsilon_F \sim t (1+\delta)$.
In PG in the BCS approximation
%, if we describe the magnetic Brillouin Zone (BZ) with the upper half of the BZ, in PG the BCS treatment of this pairing on the two small FS centered at $(\pm \pi/2, \pi/2)$ yields
 2 p-wave orders for the two holon Fermi surfaces recombine to give a d- wave order in the full BZ for the hole\cite{sus},\cite{msy1}. In SM but below $T_{ph}$
the situation is slightly more involved, but qualitatively similar \cite{mg}.
% since the pairing distinguishes the two N\'eel sublattices it produces also a folding of the holon FS into the magnetic BZ, inducing  the formation of two hole-like FS around $(\pm \pi/2, \pi/2)$ and an electron-like FS around  $(\pm \pi, 0) = (0, \pi)$. In BCS approximation we have again p-wave order in the two hole-like FS and s-wave order in the electron-like FS for the holon, reproducing finally a d-wave order for the hole in the full BZ \cite{mg}. 
The pairing turns out to be not far from the BCS-BEC crossover, but still in the BCS side with $k_F \xi_0$ ranging from to 1.5 to 3.5 for the doping region we considered, $\xi_0$ being the BCS coherence length.
However, we do not have yet condensation of holon pairs because the fluctuations of the phase of the pairing field
%In fact the scattering of the phase of the holon-pair field, with a gap $\sim T$, against holons
 destroy the holon-pair coherence of BCS approximation and produces the phenomenology of  Fermi arcs coexisting with gap in the antinodal region \cite{mg}.
Charge pairing alone is not yet hole-pairing, since the spins are still unpaired.
Only at a lower temperature $T_{ps}$ using the holon-pairs as sources of effective spin attraction, the gauge attraction between spin and charge  induces the formation of incoherent RVB spin pairs. This attraction is energetically favorable since it lowers the spinon gap from $m_s$ to $M \approx m_s-(\Delta_s)^2/m_s$ , where $\Delta_s$ is the modulus of the spinon-pair field (assumed spatially constant in MF).
 Below $T_{ps}$ the low-energy effective action obtained integrating out spinons is
% (up to  $Z_2$ vortices)
 a Maxwell-gauged 3D XY model, where the angle-field $\phi$ of the XY model is the ``square root" of the phase of the long-wave limit of the  hole-pair field $h_i^*h_j^* \epsilon^{\alpha\beta} (s_i
e^{i\Phi^s_i})_{\alpha}(e^{i\Phi^s_j}s_{j})_{\beta}$ and the gauge field is $A$. The holon contribution is QED-like but subleading. 
Introducing integer-valued vector currents $n_{\mu}$ that allow $\phi$ to be treated self-consistently as an angle function with range in the interval from 0 to  $2\pi$ (lattice regularization is understood in the following),
% and neglecting the contribution of the diluted $Z_2$ vortices \cite{note},
 the low-energy effective action $S_{\text{eff}}$ is given \cite{senthil} by:

%\begin{widetext}
%\begin{eqnarray}
%\label{eff}
%& S_{\text{eff}}
%(\phi, n_{\mu}, A_\mu)
% \approx  
% \int d^{3}x \frac{1}{6 \pi (m_s-\Delta_s^2/m_s)} [(\partial_\mu A_\nu -\partial_\nu A_\mu)^2 + \nonumber  \\ 
%& 2 \Delta_s^2 \left(\partial_{0} \phi - A_{0} + 2 \pi n_{0}\right)^2+ \Delta_s^2
%\left(\partial_{i} \phi - A_{i} + 2 \pi n_{i}\right)^2](x)\nonumber \\ &+ \frac{1}{2} (A_\mu \Pi^h_{\mu\nu} A_\nu)(x).
%\end{eqnarray}
%\end{widetext}

\begin{equation}
\begin{split}
S_{\text{eff}} (\phi, n_{\mu}, A_\mu) \approx   \int d^{3}x \Big\{ \frac{1}{6 \pi M} & [(\partial_\mu A_\nu -\partial_\nu A_\mu)^2 + 2 \Delta_s^2 \left(\partial_{0} \phi - A_{0} + 2 \pi n_{0}\right)^2+ \\
& +\Delta_s^2 \left(\partial_{i} \phi - A_{i} + 2 \pi n_{i}\right)^2] + \frac{1}{2} (A_\mu \Pi^h_{\mu\nu} A_\nu) \Big\},
\end{split}
\label{eff}
\end{equation}

where $ \Pi^h_{\mu\nu}$ is the vacuum polarization of the holons in the low-energy limit. Since in the range considered $v_F/T > v_s/T \sim J/T \gg 1$, with $v_F$ the Fermi velocity of holons and $v_s$ the spinon velocity, we extended the integration to the full Euclidean space-time $\mathbb{R}^3$, retaining though the temperature dependence of the coefficients in the action.   
%The holon contribution is QED-like but subleading. 
Neglecting the subleading contribution of holons the spinon pairing parameter $\Delta_s$ satisfies a gap equation of the form 

\begin{eqnarray}
\begin{split}
& \frac{3 }{2 m_s^2} - \frac{1}{\tau^2} = \\
& = \frac{1}{2 \left| \Delta^s_0 \right| V } \sum_\mathbf{k} \left[ \frac{k}{E_- \tanh \frac{E_-}{2T}} - \frac{k}{E_+ \tanh \frac{E_+}{2T}} \right]  
\end{split}
\label{gapeq}
\end{eqnarray}

where the energy is measured in units of $J$, $E_\pm = \sqrt{k^2+ m_s^2 \pm 2 k \Delta_s}$ and $\tau \sim |\Delta_h|$, with $\Delta_h$  the holon pairing parameter, derived in the BCS approximation.
The gauged XY model has two phases:
Coulomb and Higgs. If in (\ref{eff}) the coefficient of the Anderson-Higgs
mass term for the gauge field $A$, $\sim \Delta_s^2$, is sufficiently small,
the phase field $\phi$ fluctuates so strongly that no mass gap for
$A$ can be produced.  This
is the Coulomb phase, where a plasma of magnetic vortices-antivortices
appears. For a sufficiently
large coefficient of the mass the gauged XY  is in the broken-symmetry phase: the fluctuations
of $\phi$ are exponentially suppressed and we have $\langle e^{i \phi} \rangle \neq 0$ at $T = 0$
or there is a quasi-condensation (power-law-decaying order
parameter) at $T > 0$; accordingly magnetic vortex-antivortex
pairs become small and dilute, so the gauge field is gapped.
At the same time the holon, and hence the hole, acquires the gap outside the nodes; this is the superconducting phase.
Numerically from (\ref{gapeq}) one can check that for the critical value of $\Delta_s$
the inverse coefficient of the Maxwell term in (\ref{eff}) is quite
small, hence the gauge fluctuations are strongly suppressed.
%  and the SC transition is essentially of the XY type. 
Therefore the superconducting transition is almost of classical 3D XY type, driven by condensation of magnetic vortices and related to phase coherence as in BEC systems, in spite of the BCS-like charge pairing discussed above. This provides a solution of one of the puzzles outlined in the introduction.

%Since the coefficient of the Maxwell term is proportional to the inverse of the spinon gap, in the SC phase the gauge fluctuations are strongly suppressed.
%Numerically one can check that for the critical value of $\Delta_s$
%the inverse coefficient of the Maxwell term in (\ref{eff}) is quite
%small, hence the gauge fluctuations are strongly suppressed.
%  and the SC transition is essentially of the XY type. 
% Hence the superconducting transition is almost of classical 3D XY type, driven by condensation of magnetic vortices related to phase coherence as in BEC systems, in spite of the BCS-like charge pairing \cite{mfsy}. 

\section{The Ioffe-Larkin rule}

We now show a key property in our approach, as well as in  the slave-boson approach \cite{lee2}, of the superfluid density of the hole: it can be expressed in terms of spinon and holon contributions via Ioffe-Larkin rule \cite{il}, analogously to conductivity. To derive the superfluid density we minimally couple the holon to an external electromagnetic potential $A^{em}$, replacing $A$ by $A+ A^{em}$ in the holon contribution of eq. (\ref{eff}), denoting the result with $S_{\text{eff}}
(\phi, n_{\mu}, A_\mu ; A_\mu^{em})$, and we obtain the superfluid density as the coefficient of the quadratic Anderson-Higgs ``mass" term in the vector potential in the Euclidean  effective action for $A^{em}$ at zero energy-momenta.

Notice that the space components of $\Pi^h$ evaluated at 0 frequency-momenta give the superfluid density of the holon-gauge subsystem.
To obtain from  $S_{\text{eff}}
(\phi, n_{\mu}, A_\mu ; A_\mu^{em})$ the effective action in terms of $A^{em}$ alone we first sum over the integer currents $n_\mu$ and integrate over $\phi$. The result can be approximately summarized by replacing the XY action of eq. (\ref{eff}) by a Gaussian action for $A$, whose coefficient is the superfluid density, $\rho_s^s$, of the anisotropic 3D (Villain) XY model with effective temperature given by $\Theta \equiv 3 \pi (m_s-\Delta_s^2/m_s)/\Delta_s^2$, monotonic in $T$, with a scale renormalization $\xi$ taking into account the short-distance effects.
One can now perform the Gaussian integration over $A$ obtaining:
 \begin{eqnarray}
\label{eff1}
S_{\text{eff}}(A_\mu^{em}) \approx \frac{1}{2} \int d^{3}x \left\{ A_\mu^{em} [\frac{\rho_s^s \Pi^h }{\rho_s^s +\Pi^h }]_{\mu\nu} A_\nu^{em} \right\}
\end{eqnarray}
 The space components of $\Pi^h$ and those of the kernel of the integral in eq.(\ref{eff1}) evaluated at 0 frequency-momenta give the superfluid density $\rho_s^h$ of the holon-gauge subsystem and the physical 2D superfluid density $\rho_s$, respectively. Therefore one obtains the Ioffe-Larkin rule:
 \begin{equation}
\rho_s = \frac{\rho_s^s \rho_s^h}{\rho_s^s + \rho_s^h}.
\label{eq:sumrule}
\end{equation}

 A consequence of Ioffe-Larkin rule in this approach is the 2/3 critical exponent. In fact, $\rho_s^h$ in BCS approximation is non vanishing below the holon pairing temperature $T_{ph}$, hence it is still non-vanishing at $T_c$ where $\rho_s^s$ vanishes. Therefore $\rho_s^s$ dominates the behaviour of $\rho_s$ near the phase transition, and since $\Theta (T)$ is analytic at $T_c$, being divergent only at $T_{ps}$, the resulting critical exponent of the superfluid density is that of the 3D XY model.

The theory of the XY model suggests that
\begin{equation}
\rho_s^s(0) \approx \xi \left[\frac{\mathrm{d} \Theta}{\mathrm{d} T}(0) \right]^{-1}
\label{rhos2pre}
\end{equation}
and $\rho_s^s(T) =\rho_s^s(0) \rho_{XY}(\Theta(T)/\Theta(T_c))$, where $ \rho_{XY}$ is the spin stiffness of the anisotropic 3D XY model considered above. In the BCS approximation the holon contribution can be derived from the standard formula as

\begin{equation}
\rho_s^h(T)=\frac{2 \epsilon_F}{\pi} \left(1-\frac{\log(2)}{2 \Delta_h} T \right)
\end{equation}

in PG, taking into account the linear dispersion and the two FS of the holons.

\begin{figure}[h]
\begin{center}
{\includegraphics[trim=0 0 0 0,width=8.3cm,clip]{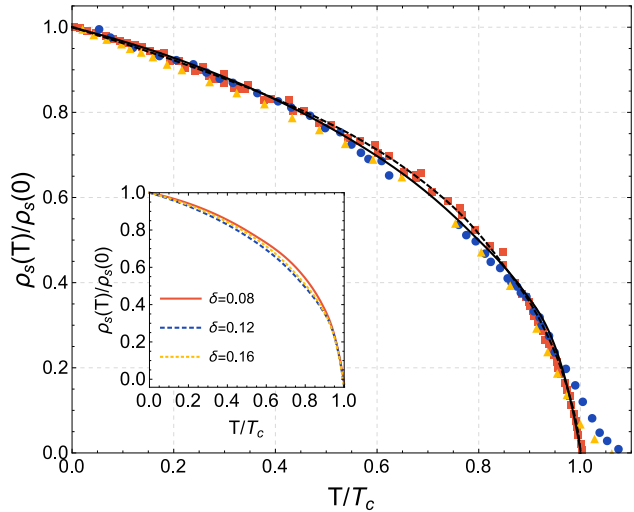}}
\end{center}
\caption{The normalized superfluid density, for $\delta=0.12$ vs. $\frac{T}{T_c}$. Our theoretical calculation (solid line, $\xi=5.6$) is compared with a pure 3D XY model (dashed line), data from \cite{hardy}, fig. 21 (squares), corresponding to the near-universal YBCO behavior for superfluid density, from \cite{panagopoulos} (circles) for $\delta=0.075$ LSCO and from \cite{jacobs}, fig 3b (triangles) for near-optimal-doping BSCCO. Data from \cite{panagopoulos} and \cite{jacobs} are plotted considering the inflection point discussed in the text. The near-universal behavior of our $\rho_s$ is shown in the inset.}
\label{fig:fig1}
\end{figure}

We now compare our results with experimental data. All the parameters (except the $\delta, T$-independent scale $\xi$) being fixed by the phase diagram as in \cite{mfsy}, we correctly fit the normalized superfluid density data for YBCO ($a$-axis) over a broad range of dopings\cite{zhang,hardy,hardy2,bonn}, see fig. \ref{fig:fig1}. 

%$b$-axis YBCO is correctly fitted only up to slight underdopings; at optimal doping and beyond the effect of the chains, and the consequent YBCO $a-b$ anisotropy, ruins the agreement. In \cite{jacobs} BSSCO data is shown to be be very compatible with $a$-axis YBCO data in \cite{zhang}; we fit well both. LSCO data in \cite{panagopoulos} is well fitted for the underdoped samples, except for a small region near $T_c$; however the authors in \cite{panagopoulos} report that doping inhomogeneity may affect the reliability of experimental data in this very region. Similarly, Hg-1201 data (\cite{panagopoulos}) is also fitted reasonably well in the underdoped regime.
Analogous data in moderate underdoped BSCCO \cite{jacobs}, Hg-1201 \cite{panagopoulos} and LSCO \cite{panagopoulos} are well fitted except for a small region near $T_c$, beyond an inflection point in the curve, where, however, the authors of \cite{panagopoulos} claim that doping inhomogeneities may affect the reliability of the experimental data.

\section{The Uemura relation}

\begin{figure}[h]
\begin{center}
{\includegraphics[trim=10 10 0 20,width=8.3cm,clip]{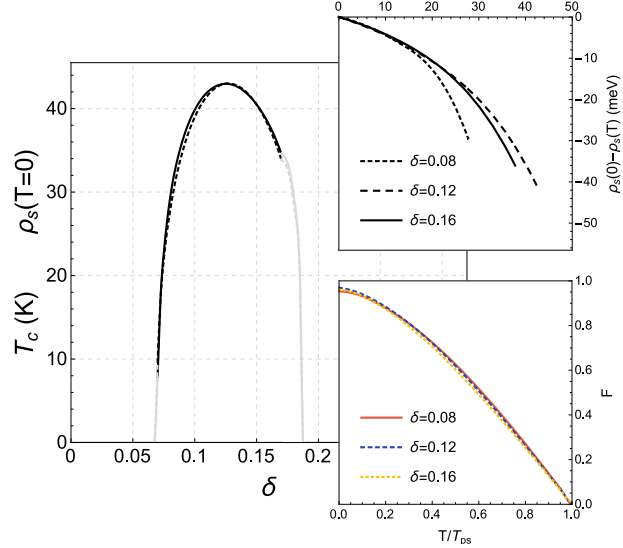}}
\end{center}
\caption{Theoretically calculated $T=0$ superfluid density (solid line, arbitrary units) and critical temperature (dashed line) vs. $\delta$ exhibiting an approximate Uemura relation over the broad doping range considered here. Insets: approximate $\delta$-universality of the slope of $\rho_s$ near $T=0$ (upper inset) and  of $F$ (lower inset).}
\label{fig:fig2}
\end{figure}

We now prove analytically that from the temperature dependence of $\Delta_s$ follows an approximate Uemura relation. Let us suppose, as the numerics proves, that the spinon contribution dominates the superfluid density also at $T=0$. Since $\Delta_s$ is the solution of a gap equation with critical temperature $T_{ps}$, its temperature dependence will be a function of $T/T_{ps}$. Furthermore one can check numerically that $T_c \ll T_{ps}$ and for a wide range of dopings, more precisely slightly away from the doping at which $T_{ps}$ vanishes, $\Delta_s^2 (T=0) \approx m_s^2$.
% hence the effective temperature at $T=0$ of the XY model in this doping range is approximately 0. 
Therefore we can parametrize   $\Delta_s^2 (T) = m_s^2 F(T/T_{ps})$ for a suitable function $F$ obtained from (\ref{gapeq}), which turns out to be almost independent of $\delta$ and with $F(0) \approx 1$  in the above quoted range of dopings, see fig. \ref{fig:fig2}, lower inset.
The effective temperature $\Theta$ defined in the previous section can then be written in terms of $F$ as:

$$ \Theta(T) = \frac{3 \pi}{m_s} \frac{1-F(T/T_{ps})}{F(T/T_{ps})} $$

so that the (dominating) spinon contribution to superfluid density in eq. (\ref{rhos2pre}) can also be written as:

\begin{equation}
\rho_s^s(0) \approx \frac{ \xi m_s T_{ps} F(0)^2}{3 \pi |F'(0)|} .
\label{rhos2}
\end{equation}

%The holon contribution can be derived from the standard formula for fermions as $\rho_s^h(T)=\epsilon_F/(4 \pi)(1- \gamma \log[2] T/ \Delta_h)$ and it turn out to be a monotonic increasing function of doping. $\gamma$ is our only free parameter, parametrizing possible corrections to the BCS approximation and is fixed by comparison with experiments at one doping, and then assumed doping independent; the final results are only weakly dependent on it.

If we impose the condition of criticality to the effective temperature of the XY model: $\Theta(T_c) \equiv T^{XY}_c \approx 2.2$ \cite{tc}, expanding $F$ linearly, which can actually be justified numerically, we obtain:

\begin{equation}
\label{Tc}
T_c \approx  \frac{T_{ps}}{|F'(0)|} \frac{F(0)-1 + F(0) m_s  T^{XY}_c/(3 \pi)}{1+ m_s   T^{XY}_c/(3 \pi)}
\end{equation}

and the derived shape of $T_c(\delta)$ in PG is approximately parabolic, as in the phenomenological curve first proposed in \cite{presland}, see fig. \ref{fig:fig2}.

In the region where $F(0) \approx 1$ comparing with (\ref{rhos2}) and using  Ioffe-Larkin rule one obtains:

\begin{equation}
\label{uemura}
T_c \approx  \frac{T^{XY}_c}{\xi} \rho_s(0) \frac{1}{(1- \rho_s(0) / \rho_s^h(0))(1+ m_s  T^{XY}_c/(3 \pi))}
\end{equation}

Since $ m_s T_c^{XY} / (3 \pi) \ll 1$  and the spinon contribution is dominating,  eq. (\ref{uemura}) yields the Uemura relation, see fig. \ref{fig:fig2}.
%, that actually occur if in the doping range considered there is a region where $T_c$ is non monotonic in $\delta$ whereas $\rho_s(0)$ is monotonic increasing. 

%\begin{figure}[h]
%\begin{center}
%{\includegraphics[trim=0 0 0 0,width=8.3cm,clip]{p3.eps}}
%\end{center}
%\caption{Universality of $\rho_s (T) - \rho_s (0)$ vs. $T$ for different $\delta$ values, as observed in experimental data in \cite{staj}: this feature is reproduced quite accurately by our theoretical model. Units of $J$ are used on both axes.}
%\label{fig:fig3}
%\end{figure}

Last but not least, since $\Theta(T)$ is approximately proportional to $T$, the ratio $\Theta(T) / \Theta(T_c) \approx T/T_c$, hence when the holon contribution is small the superfluid density $\rho_s(T)/\rho_s(0)$ is approximately the universal function of $\rho_{XY}(T/T_c)$ derived from the 3D classical XY model in the entire temperature range, not only close to the phase transition, and this property is quite well verified by the experimental data, as fig. \ref{fig:fig1} shows.
From the above property and the Uemura relation one derives also $\rho_s(0) \approx c T_c$ with $c$ doping-independent and expanding $\rho_s(T)/\rho_s(0)$  linearly in $T/T_c$, we can write $\rho_s(T) - \rho_s(0) \approx c (\mathrm{d} \rho_{XY}/ \mathrm{d}(T/T_c))(0) T$. Hence in the considered doping region the $T$-slope of the superfluid density near $T=0$ is almost doping independent, see fig. \ref{fig:fig2}, as discussed experimentally in \cite{staj}.

\section{Conclusions}
Summarizing, we have shown that many experimental features of the superfluid density of hole-doped cuprates in the intermediate doping region are captured by an approach of the hole as a gauge composite of charge and spin excitations. With a single $(\delta, T)$-independent scale parameter $\xi$  keeping track phenomenologically of small scale physics, our model fits rather well normalized superfluid density data from YBCO ($a$-axis), BSCCO, Hg-1201 and LSCO from moderate underdoping to nearly optimal doping. The universal critical exponent $2/3$ and the near-universality of 3D XY type of the normalized superfluid density are also reproduced independently of $\xi$, as well as the approximate Uemura relation which is analytically derived. 

These results suggest that the spin-charge composite nature of the hole plays a key role in high-$T_c$ superconductivity.

% The effect of magnetic impurities and of the magnetic resonance in this approach to superconductivity will be analyzed elsewhere.

%A more detailed account of the approach to the superconducting phase discussed here, including the effect of magnetic impurities and the analysis of the magnetic resonance will be presented elsewhere.
%We end with a brief comment about the superfluid density in the doping regions not considered here.
%At very low doping our approach can not be safely trusted because theoretically the corresponding normal state region is not correctly described by our treatment of the PG phase, presumably from the experimental point of view because it does not take into account the "spin-glass" phenomenology. 
%%However, for very small doping one can anticipate a dominant contribution of holons except near $T_c$, thus leading to an essentially linear behaviour of $\rho_s(T)$, which is still not inconsistent with the data.
% At large doping we believe that holon and spinon becomes more tightly bound, producing in the superconducting phase a more standard BCS-like behaviour, but we have not yet worked out the details.

\section*{Acknowledgments}

Partial support of Cariparo
Foundation (Excellence Project ``Macroscopic Quantum Properties of Ultracold
Atoms under Optical Confinement"), and of MIUR (PRIN Project ``Collective Quantum Phenomena: from Strongly-Correlated Systems to Quantum Simulators") are gratefully acknowledged.
P.A.M. thanks Z.B. Su, F. Ye and L. Yu for useful discussions.

\end{document}